 \documentclass[twocolumn,aps,prl,showpacs,superscriptaddress]{revtex4}

\usepackage{graphicx}
\usepackage{dcolumn}
\usepackage{bm}

\def\duzomniejsze{<\kern-.7mm<}
\def\duzowieksze{>\kern-.7mm>}

\def\textbf#1{{\bf #1}}
\def\beq{\begin{equation}}
\def\eeq{\end{equation}}
\def\be{\begin{equation}}
\def\ee{\end{equation}}
\def\ben{\begin{eqnarray}}
\def\een{\end{eqnarray}}
\def\beqa{\begin{eqnarray}}
\def\eeqa{\end{eqnarray}}
\def\eea{\end{array}}
\def\bea{\begin{array}}
\newcommand{\bei}{\begin{itemize}}
\newcommand{\eei}{\end{itemize}}
\newcommand{\bee}{\begin{enumerate}}
\newcommand{\eee}{\end{enumerate}}

\begin{document}


\title{Intrinsic asymmetry with respect to adversary:
new feature of Bell inequalities}

\author{Pawe{\l} Horodecki}

\affiliation{Faculty of Applied Physics and Mathematics, Gda\'nsk University of Technology, 80-233 Gda\'nsk, Poland}
\affiliation{National Quantum Information Centre in Gda\'nsk, 81-824 Sopot, Poland}

\author{Marcin Paw{\l}owski}

\affiliation{Institute of Theoretical Physics and Astrophysics, University of Gda\'nsk, 80-952 Gda\'nsk, Poland}
\affiliation{School of Mathematics, University of Bristol, Bristol BS8 1TW,
United Kingdom}

\author{Ryszard Horodecki}
\affiliation{Institute of Theoretical Physics and Astrophysics, University of Gda\'nsk, 80-952 Gda\'nsk, Poland}
\affiliation{National Quantum Information Centre in Gda\'nsk, 81-824 Sopot, Poland}

\date{\today}
\begin{abstract}
It is known that the local bound of a Bell inequality is sensitive
to the knowledge of the external observer about the settings
statistics. Here we ask how that sensitivity depends on
the structure of that knowledge. It turns
out that in some cases it may happen that the
local bound is  much more sensitive to adversary's knowledge
about settings of one party than the other.
Remarkably, there are Bell inequalities which are highly asymmetric
with respect to the adversary's knowledge about local settings.
This property may be viewed as a hidden
intrinsic asymmetry of Bell inequalities. Potential implications of the
revealed asymmetry effect are also discussed.
\end{abstract}

\maketitle

\section{ Introduction}
Bell inequalities \cite{Bell-review} are the unique tool to prove
the difference between quantum and classical world both
in the philosophical and practical way.
In fact, it was the famous Einstein-Rosen-Podolsky
\cite{EPR} paper which introduced the concept of local realism. Yet, no method of experimental verification of
their claims was given there. It was only found by Bell in his famous paper
\cite{Bell}.  While it was clear that entanglement
\cite{Schroedinger} which lead to the paradox, Bell was the first
to propose the inequality that under natural
assumptions (i) spatial separation of two systems and (ii) locally realistic description
of the results of the measurements (that may be thought of
as objective properties of two systems) must be necessarily
satisfied. Quantum mechanics however violates the
prediction of many Bell inequalities (see however
\cite{GYNI}).

In fact any
violation of Bell inequality leads to refutation of locally
realistic theory \cite{Bell-review}. On the other hand
it is a unique tool for device independent quantum cryptography
\cite{device-independent-crypto}, and randomness amplification against
quantum \cite{MironowiczPawlowski} even non-signaling
adversary \cite{CollbeckRenner-NP12,Gallego-NC13}.

However there is a natural boot-strap problem, which is a
free randomness assumption \cite{cytat-free-will}
which correspond to the duality: to interpret violation
of any Bell inequality bound as a signature of
absence of local realism one must ensure that
the data come form the experiment where
the statistics of the settings were intrinsic random, i.e.
they should have been uncorrelated from everything else in the world,
which is represented as a statistical independence of
the source itself with respect to the
rest of the universe, or in  other words "statistical unpredictability"
to the rest of the universe.
This  is why sometimes this randomness is called "the observer's
free will". Recently the effects  of reduced "free will"
on Bell inequalities have been demonstrated in different contexts \cite{Hall-free-will,BarrettGisin,CollbeckRenner-NP12,Brukner,Hall-relaxed-Bell}.
In particular it has been showed that the knowledge of the adversary about the settings
statistics can have dramatic consequences on the usual interpretation
of the Bell inequalities \cite{Hall-free-will,cytat-free-will}. In this paper we pose a more sophisticated  question:{\it How
the structure of that knowledge influences the local bound of a Bell inequality?}
It turns out that answer to this question lies in the structure of the Bell inequalities itself.
More precisely we show that there are Bell inequalities highly asymmetric with respect to
the adversary knowledge about local settings.
\section{ No-signaling boxes and Bell inequality}
Consider any given 2-party no-signaling box \cite{PopescuRohlich}
which is represented by the family of conditional probability
distributions
\begin{equation}
{\bf P}:=\{ p(ab|xy)\},
\label{box}
\end{equation}
with settings $x \in \{1,...,N_{A}\}:=I_X$, $y \in \{ 1,...,N_{B}\}:=I_{Y}$
and outcomes  $a \in \{1,...,M_{A}\}:=I_A$, $b \in \{ 1,...,M_{B}\}:=I_B$.
We call the box no-signalling form Alice to
Bob (from Bob to Alice) if, as usual, $\sum_a p(ab|xy)=p(b|y)$
($\sum_b p(ab|xy)=p(a|x)$), i.e. local statistics of one party
does not depend on the settings of the other party.
Suppose that the box ${\bf P}_{LHV}:= \{ p_{LHV}(ab|xy)\}$ satisfies the Local Hidden Variable (LHV) description,
namely there are the following conditional probabilities
$\{ p(a|x,\lambda)\}$, $\{ p(b|y,\lambda)\}$ and some
probability distribution on some probabilistic space
$\Lambda$,  $\lambda \in \Lambda$ such that
\begin{equation}
p_{LHV}(ab|xy)= \sum_\lambda p(\lambda) p(a|x,\lambda) p(b|y,\lambda).
\label{box-LHV}
\end{equation}
Note that in the case of Bell inequalities with a finite number of settings, there is also only a finite number of pure classical strategies (and every other strategy is a mixture of them). This enables us to restrict ourselves to a finite alphabet of $\lambda$'s, which describe the choice of the strategy. Therefore, in our paper we can use sums instead of integrals for the description of classical strategies.

Let us formulate the Bell observable value as follows
\begin{equation}
B({\bf P})=\sum_{a,b,x,y} p(ab|xy) B(a,b,x,y) \alpha(a,b,x,y) p(x,y)
\label{Bell-value}
\end{equation}
In the above $B(a,b,x,y)$ is an indicator function
such that $ B(a,b,x,y)= 1 \Leftrightarrow (a,b,x,y) \in {\bf \Omega} \subset {\bf I}:=I_A \times  I_B
\times I_X \times  I_Y$ and $ B(a,b,x,y)= 0$ otherwise.
The probabilities ${ p(x,y) }$, satisfying $\sum_{x,y}p(x,y)=1$
represent the probabilities of the settings of the inequality.
The conditional probabilities describe the behaviour of the box while
$\alpha(a,b,x,y)$ describes the {\it pay-off function}.
Now the Bell inequality may be considered to be any inequality of the form:
\begin{equation}
B({\bf P}_{LHV}) \leq R_{LHV}
\label{Bell-inequality}
\end{equation}
giving the bound for all Bell observable values achievable by boxes satisfying LHV theories.

We use the language of game theory here because it is often more convenient to treat Bell inequalities as nonlocal games. In this approach the box plays against a referee, the provides the settings according to the distribution ${ p(x,y) }$. The conditional probabilities describe the strategy of the box and the pay-off function the winnings in each case. This treatment of Bell inequalities is especially useful if there are additional constraints involved in the problem at hand, e.g. while preparing the strategy the box is given the distribution of the settings ${ p(x,y) }$ only approximately, or one of the players apart from learning his or her input learns something about the input of the other party. The second case is exactly what we analyze in this paper.

Before we proceed, we should stress that for every Bell inequality there are many inequivalent nonlocal games. This is easily seen in formula (\ref{Bell-value}) where the same value of the product $\alpha(a,b,x,y) p(x,y)$ can be obtained with many different combinations of factors. Therefore, every nonlocal game should be considered a {\it representation} of a Bell inequality rather than the inequality itself. However, in any experiment, an inequality is not tested but a nonlocal game is played (a probability distribution of the settings must be clearly defined and pay-off function applied in the data processing phase).

\section{Bit rate of Eve's knowledge}
Consider now a specific Bell inequality in which the
statistics of the settings are independent (usually also assumed to be uniform, but they do not need to be):
\begin{equation}
p(x,y)=p(x)p(y)
\label{uniform}
\end{equation}

Now consider an adversary - Eve - who has access to some two hidden
parameters $\lambda_1, \lambda_2$ correlated to the settings, namely they have at their
 disposal some conditional statistics:
\begin{eqnarray}
&& p(x|\lambda_1),\  \sum_{\lambda_1} p(x|\lambda_1)=p(x)
\nonumber \\
&& p(y|\lambda_2), \ \sum_{\lambda_2} p(y|\lambda_2)=p(y)
\label{controlled-settings}
\end{eqnarray}
with some specific ''hidden variables'' $\lambda_i \in \Lambda_i$, $i=1,2$ representing
the Eve knowledge about each of the settings.
The summation conditions in the above represent the fact that neither
Alice nor Bob is supposed to notice any change of the statistics
of their settings despite there is a conditional control of
them by external adversary.

Now we shall consider the situation when there is specific
knowledge of Eve about either  one of the settings. Consider any
fixed measure of entropy $H$. For now we do not specify the particular
form of that entropy function.
Obviously for any specific statistics of the local settings
$\{ p(x) \}$ and $ \{ p(y) \} $ we have the
corresponding entropies of {\it local statistics of settings} $H(X)$ and $H(Y)$.
The knowledge of Eve about the statistics is described by
the  {\it conditional entropies of the settings}  (defined consistently
with the entropy $H$ above, whatever it was chosen to be) which
we shall denote as $H(X|\Lambda_1)$, $H(Y|\Lambda_2)$.

Now we shall introduce the notion of {\it relative knowledge} of Eve
about the statistics of local settings (cf. application in  standard
communication scheme  \cite{Walczak}) following \cite{CT}:
\begin{eqnarray}
&& \xi_{X}=\frac{H(X)- H(X|\Lambda_1)}{H(X)} \nonumber \\
&& \xi_{Y}=\frac{H(Y)- H(Y|\Lambda_2)}{H(Y)}
\label{percentage}
\end{eqnarray}
which in fact represents the {\it bit rate} of the Eve's knowledge
about the settings which describes how big is the ratio of
the total randomness of the local settings to the apparent.
Here
$\xi_{X},\xi_{Y} \in [0,1]$ and the case
of, say $\xi_{X}=0$, ($\xi_{X}=1$) corresponds to zero
and maximal Eve's knowledge respectively.

We choose this measure of information because it is invulnerable to differences in the number of settings per party. For example consider the family of inequalities introduced in \cite{M-EARAC}. There one party has exponentially more settings than the other. Therefore, e.g. 10 bits of information can be at the same enough to fully specify the setting of one party while only being able to encode 0.1\% of information about the setting of the other.

\section{ Defining the intrinsic asymmetry of Bell inequality}
Now the central quantity of this paper is the new Bell value
\begin{equation}
\overline{B}({\bf \tilde{P}}_{LHV,\xi_{X},\xi_{Y}})\leq \tilde{R}_{LHV}(\xi_{X},\xi_{Y})
\label{Bell-new}
\end{equation}
The quantity on the right hand side value reproduces the standard Bell bound for
complete lack of knowledge of Eve ie.
\begin{equation}
R_{LHV}=\tilde{R}_{LHV}(\xi_{X},\xi_{Y}),
\end{equation}
which is the minimal value of $\tilde{R}_{LHV}(0,0)$.

The left hand side $\overline{B}({\bf \tilde{P}}_{LHV,\xi_1,\xi_2})$ is any  value calculated for a LHV boxes strategy, ie. the family
of correlated LHV boxes prepared by Eve
\begin{equation}
\tilde{p}_{LHV}(ab|xy\lambda_1\lambda_2)= p(a|x,\lambda_1) p(b|y,\lambda_2)
\label{box-LHV1}
\end{equation}
Inserting it to (\ref{Bell-value}) we get
\begin{eqnarray}
&& \overline{B}({\bf P}_{LHV,\xi_{X},\xi_{Y}})= \nonumber \\
&& \sum_{\lambda_1,\lambda_2} p(\lambda_1,\lambda_2)\sum_{a,b,x,y}  p(ab|xy\lambda_1 \lambda_2) \nonumber \\
&&  B(a,b,x,y) \alpha(a,b,x,y) \nonumber \\
&& p(x|\lambda_1) p(y|\lambda_2)
\label{Bell-value-new}
\end{eqnarray}
that is on the left hand side of the {\it Eve's knowledge dependent
Bell inequality} (\ref{Bell-new}). The bar over the quantity $B$ stresses the fact that
this represents the mean value of the different ($\lambda_1,\lambda_2$ - dependent)
averages in the experiment.
Actually RHS of that inequality, which is a new local realistic
bound $\tilde{R}(\xi_X,\xi_Y)$ can be seen as a maximum of the LHS over all families of the boxes
$\{ \tilde{p}_{LHV}(ab|xy \lambda_1 \lambda_2)\}$ and
the associated probability distributions (\ref{controlled-settings})
such that the corresponding entropies of the settings of the original inequality
and the present inequality satisfy the fixed percentage conditions (\ref{percentage}).

Note that from the above formulae one can
naturally construct the sensitivity indicators of the increase of the Bell value under
existence of extra information  about the settings in the adversary's hands:
$D_{A}(\xi_{1}|H(X)) = R_{LHV}(\xi_1,0)-R_{LHV}(0,0),  D_{B}(\xi_{2}|H(Y)) = R_{LHV}(0,\xi_2)- R_{LHV}(0,0)$.

\section{Searching for hidden asymmetry in Bell inequality}

From now on we shall consider the inequalities
that originally involve maximally mixed distribution
of settings i.e.
\begin{eqnarray}
&&p(x)=\frac{1}{N_{A}} \nonumber \\
&&p(y)=\frac{1}{N_{B}}
\label{uniformity}
\end{eqnarray}
Then {\it any difference}
between $D_{A}(\xi_{X}=\xi|\log N_{A})$, and the  $D_{B}(\xi_{Y}=\xi|\log  N_{B})$
is an indicator of the
intrinsic asymmetry of the investigated Bell inequality.
We shall introduce below the quantity that will reproduce that
difference as a special case.

In the analysis below we take a specific entropy measure, namely the so called {\it min-entropy}
\begin{equation}
H(X)=-\min_{x \in \cal{X}}\log p(x)
\label{min-entropy}
\end{equation}
and {\it  conditional min-entropy}:
\begin{equation}
H(X|\Lambda)=-\sum_{\lambda\in  \Omega_{\Lambda}} p(\lambda) \min_{x \in \cal{X}}\log  p(x|\lambda)
\label{min-entropy}
\end{equation}
where  $\Omega_{\Lambda}$ is the probabilistic space of the random variable $\Lambda$.
\subsection{Asymmetry indicators}
For fixed values of the standard statistics in the Bell value
there are several possibilities to provide the asymmetry indicators.
First, given the values  $\tilde{R}_{LHV}(\xi_{X}, 0)$,
$\tilde{R}_{LHV}(0,\xi_{Y})$ one may define
a quantity
\begin{equation}
\Delta(\xi)=|\tilde{R}_{LHV}(\xi, 0) -  \tilde{R}_{LHV}(0,\xi)|,
\label{1-parametrowy-indykator}
\end{equation}
by putting  $ \xi_{X}=\xi_{Y}=\xi$.
There is however yet more general option, namely one can depict the fully symmetric quantity
\begin{equation}
\Delta(\xi_X,\xi_Y)=|\tilde{R}_{LHV}(\xi_X, \xi_Y) -  \tilde{R}_{LHV}(\xi_Y,\xi_X)|
\label{symmetric-indicator}
\end{equation}
Note that all the above quantities (\ref{1-parametrowy-indykator}) and
(\ref{symmetric-indicator}) are calculated for some fixed values of a-priori
fixed entropies $H(X)$,$H(Y)$.
It is also good to remember that $\Delta(\xi)=\Delta(\xi_X=\xi,\xi_Y=0)$.

\subsection{Checking some table-encoded inequalities}
For our purposes it is convenient to present Bell inequalities as tables that specify the product $ B(a,b,x,y) \alpha(a,b,x,y)$. For example, the well known CHSH inequality is
\be
\begin{tabular}{c|c|c|c|c}
      & x,a=0,0 & 0,1 & 1,0 & 1,1  \\
 \hline
 y,b=0,0 & 1  & -1  & 1 & -1  \\
 \hline
 y,b=0,1 & -1 &  1  & -1 & 1 \\
 \hline
 y,b=1,0 & 1  &  -1 & -1 & 1 \\
 \hline
 y,b=1,1 & -1 & 1 & 1 & -1 \\
 \hline
\end{tabular}
\ee
This inequality is invariant under the permutation of the parties, which is reflected here by the invariance of the table under transposition. Any inequality with this property has $\tilde{R}_{LHV}(\xi_X, \xi_Y)=  \tilde{R}_{LHV}(\xi_Y,\xi_X)$ and displays no asymmetry.

\section{Quantizing the asymmetry}

However, there are inequalities without this inherent symmetry.
For example, take $I_{3322}$ \cite{GisinCollins} which is described by the table:
\be
\begin{tabular}{c|c|c|c|c|c|c|c|c}
      & x,a=0,0 & 0,1 & 1,0 & 1,1 & 2,0 & 2,1 & 3,0 & 3,1 \\
 \hline
 y,b=0,0 & 0  & 0  & 0 & 1 & 0 & 0 & 0 & 0  \\
 \hline
 y,b=0,1 & 0  & 0  & 0 & 1 & 0 & 0 & 0 & 0  \\
 \hline
 y,b=1,0 & 0  & 0  & 1 & 0 & 1 & 0 & 1 & 0  \\
 \hline
 y,b=1,1 & 2  & 2  & 0 & 0 & 0 & 0 & 0 & 0  \\
 \hline
 y,b=2,0 & 0  & 0  & 1 & 0 & 1 & 0 & 0 & 1  \\
 \hline
 y,b=2,1 & 1  & 1  & 0 & 0 & 0 & 0 & 1 & 1  \\
 \hline
 y,b=3,0 & 0  & 0  & 1 & 0 & 0 & 1 & 0 & 0  \\
 \hline
 y,b=3,1 & 0  & 0  & 0 & 0 & 1 & 1 & 0 & 0  \\
 \hline
\end{tabular}
\ee
This in one of many equivalent representations of $I_{3322}$ in which no negative values appear. Moreover a fourth setting ($x=0,y=0$) is added for each party which corresponds to measuring marginal probabilities in the original version. This choice is quite natural. It comes from a problem of obtaining the marginal probability distribution from experimental data where only correlated events are recorded. One could compute it by summing all the events when other party chose a particular setting. Usually, no-signalling principle would guarantee that the value would be the same regardless of the choice of the other party's setting. However, in our case no-signalling does not apply - the setting of one party can be transmitted to the other via the source.

We plot $\Delta(\xi)$ for this inequality in Fig. 1.

\begin{figure}[!h]\center
\resizebox{9cm}{!}{
\includegraphics{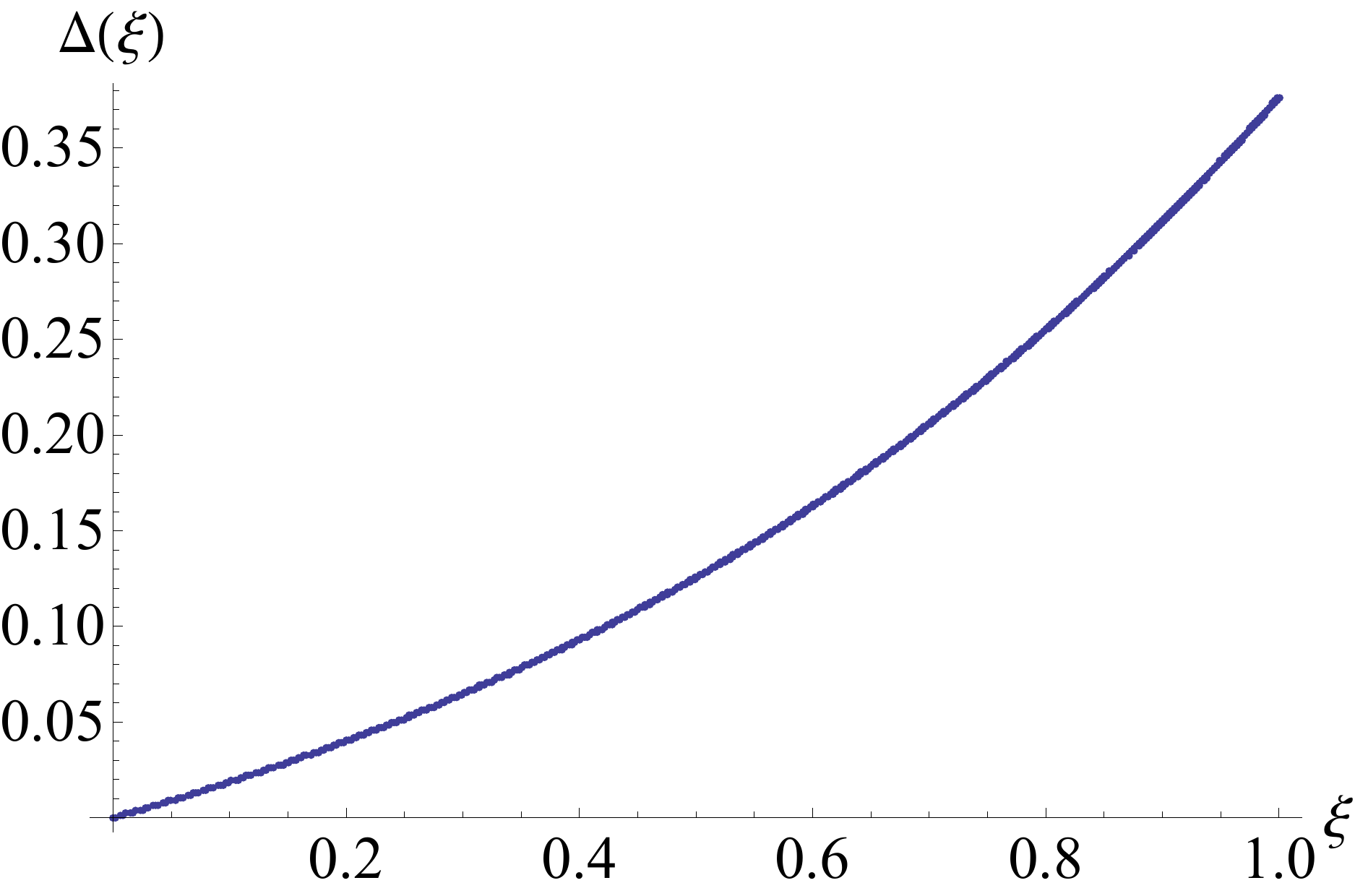}}
\caption{Asymmetry of $I_{3322}$ inequality: Function $\Delta(\xi)$.}
\end{figure}

\section{Conclusions and discussion}
We have identified a hidden intrinsic property of Bell inequalities
which is sensitive to the structure of adversary's knowledge about local settings.
We have introduced the parameter showing how
given Bell inequalities are sensitive to the
structure of Eve's knowledge about the setting of the inequality.
The same percentage of the information prompted by Eve
to the experimentalists about left and right settings can lead to
less or more fake values $\tilde{R}_{LHV}(\xi_{X}, 0)$,
$\tilde{R}_{LHV}(0,\xi_{Y})$ in
the sense that they exceed the standard LVH bound
known as a Bell inequality.
Their difference $\Delta=|\tilde{R}_{LHV}(\xi_{X}, 0) -  \tilde{R}_{LHV}(0,\xi_{Y})|$ is
a natural indicator of the hidden asymmetry of sensitivity of
the inequality to the leakage of knowledge to external
adversary.
There are several possibilities how this work can be generalised.
First we may {\it drop the uniformity assumption}
(\ref{uniformity}) as a reference point
in calculating the parameters
$\xi_X$, $\xi_Y$.
One might see the interesting interplay since
the same percentage may either {\it become more or less important}
for the case when the reference statistics is no longer
uniform.

Another natural extension would be  the multipartite scenarios,
where an interesting possibility of analysis might be to consider
the local and partially nonlocal knowledge of the adversaries
about the settings statistics.

As for practical implications the general sensitivity of
the Bell inequality value on the process of a-priori prompting the
setting  information by Eve
may be very important in the case of untrusted devices
especially in the case of high pay-off losses in device independent cryptography.
Given two or more location it is always good to put
the most robust part of the scheme in the lab that may
be most sensitive to prior prompting of the preexisting
value by the external adversary to the observers,
whatever mechanism it would be.

Also, if postprocessing of the raw data is applied and rounds when at least one of the detectors did not register a particle discarded, then we need to either assume "fair sampling" or face an important problem of experimental Bell inequality violations: detection efficiency loophole. In this case local hidden variables, by controlling when the detectors do not "click" can introduce correlations between the source of the states and settings in the data left after discarding. If an experimentalist has detectors with (even slightly) different efficiencies then they will be correlated with the source with different strengths. Our results can help experimentalists to choose which party should be granted the better detector.

There is a fundamental open question what is the potential importance
of the revealed hidden asymmetry of the sensitivity for
eavesdropper prompting the settings from
the perspective of foundations of information theory and
foundations of physics.

There is an important issue here. As already mentioned in the introduction
there are many mathematically equivalent forms of a given Bell inequality
that follow from the normalisation of the probabilities involved.
In a particular experiment only one of them corresponding to the
specific nonlocal game is tested.

At that moment it is an open question whether the asymmetry revealed
in the present paper is the feature of the particular form
of the Bell inequality (equivalently - specific nonlocal game)
or they concern all the Bell inequalities that are mathematically
equivalent to each other. This issue is left for further research.
No matter which of the two variants is true, we believe that
the value of the present result is important. Indeed
(i) either all the Bell inequalities within the equivalence class
are asymmetric - may be to the different degree - in the present sense -
and then we deal with some novel
ontological feature of Bell inequality itself or (ii) it  is a property of
some representatives of the class, which would mean that in fact
that its members were only {\it apparently equivalent}.

We conjecture that the first part of the alternative is true, ie.
we expect that any representative (game) of $I_{3322}$, which table is not invariant under the permutation of the parties\footnote{Invariant representatives of $I_{3322}$ are known to exist\cite{InvI}.}, to exhibit the asymmetry, but to different degree. This would
imply both "ontological" (independent on the implementation) meaning of
the observed asymmetry as well as novel practical implication:
we should rather talk about and work with particular nonlocal
games rather formal Bell inequalities.
Shortly speaking this would mean that what was believed to be a single
equivalence class is split due to practical reasons that must not be ignored.
We hope that the answer to the above questions opened by the present
result will be found soon.

\section{acknowledgements}

This paper is supported by the
Polish Ministry of Science and Higher Education
Grant IdP2011 000361, NCN grant 2013/08/M/ST2/00626, FNP TEAM programme and ERC grant QOLAPS.
Part of this work was done when P. Horodecki and  M. Paw{\l}owski
were  attending
the Programme ,,Mathematical Challenges in Quantum Information
Isaac Newton Institute for  Mathematical Sciences in Cambridge.


\end{document}